\documentclass[preprint,aps]{revtex4}
\usepackage{graphicx}  
\usepackage{bm}
\usepackage{epsfig}
\usepackage{color}
\newcommand{\be}{
\begin{equation}
}
\newcommand{\ee}{
\end{equation}
}
\newcommand{\ba}{
\begin{eqnarray}
}
\newcommand{\ea}{
\end{eqnarray}
}
\newcommand{\vc}[1]{
{\bf{#1}}
}

\newcommand{\pard}[2]{
{\frac{\partial #1}{\partial #2}}
}

\newcommand{\df}{
{\rm{d}}
}
\newcommand{\wt}[1]{
\widetilde{#1}
}
\newcommand{\gav}[1]{
\langle #1 \rangle
}
\newcommand{\Bgav}[1]{
\Big \langle #1 \Big \rangle
}
\newcommand{\intl}{
\int \limits
}

 \newcommand{\graph}[2]{
 \newpage
 \begin{figure}[hhh]
 \begin{minipage}{16.cm}
 \epsfclipon
 \epsfysize 16.truecm 
 \psfig{figure=#1.eps,width=7.62cm}
 \end{minipage} \\
 \caption{(Color online) #2}
 \label{#1} 
 \end{figure}
 }

\begin{document}
%
\title{Pullback transformation in gyrokinetic electromagnetic simulations}
%
\author{Alexey Mishchenko\footnote[1]{alexey.mishchenko@ipp.mpg.de}}
\affiliation{Max Planck Institute for Plasma Physics,
  D-17491 Greifswald, Germany}
\author{Axel K\"onies}
\affiliation{Max Planck Institute for Plasma Physics,
  D-17491 Greifswald, Germany\quad}
\author{Ralf Kleiber}
\affiliation{Max Planck Institute for Plasma Physics,
  D-17491 Greifswald, Germany\quad}
\author{Michael Cole}
\affiliation{Max Planck Institute for Plasma Physics,
  D-17491 Greifswald, Germany\quad}
%
%
\date{\today}
%
%
\begin{abstract}
It is shown that a considerable improvement in the global gyrokinetic
electromagnetic simulations can be achieved by a slight modification of the
simulation scheme. The new scheme is verified, simulating a Toroidal Alfv\'en
Eigenmode in tokamak geometry  at low perpendicular
mode numbers, the so-called ``MHD limit''. Also, an electromagnetic drift mode
has been successfully simulated in a stellarator.
%
%
\end{abstract}
%
\maketitle
%
%
\newpage
%
%
%
\section{Introduction}
Electromagnetic effects, such as Alfv\'en waves or tearing dynamics, 
are of importance in fusion plasmas. A major complication for such simulations
is caused by the so-called cancellation problem
\cite{Chen_Parker_2001,Mishchenko1}. This problem has been 
addressed by various authors both in the particle-in-cell (PIC)
\cite{Chen_Parker_2003,Mishchenko1,Mishchenko2,Hatzky_2007} and Eulerian \cite{GYRO}
numerical framework. The scheme described in
Refs.~\cite{Chen_Parker_2003,Mishchenko1,Mishchenko2,Hatzky_2007,GYRO} is
used by a number of numerical codes, but its performance may be inhibited,
particularly when performing global simulations in realistic shaped geometries
and at realistic plasma $\beta$ values. Recently, an approach
\cite{Mishchenko_MHD} has been suggested, based 
on a novel choice of the gyrokinetic variables, which makes it possible to
mitigate the cancellation problem, in particular for Alfv\'enic-type dynamics
where $E_{\|} \approx 0$. The drawback of this method is that it is
limited to cases in which Alfv\'enic dynamics dominates. It does not help 
in the electromagnetic drift mode simulations and its performance
diminishes considerably when diamagnetic drift effects become of importance,
e.g.~for the drift-kink instabilities. In this 
paper, we describe an algorithm 
overcoming such limitations. 

We verify our scheme by simulating the 
Toroidal Alfv\'en Eigenmode (TAE) in tokamak geometry and the electromagnetic
Ion Temperature Gradient-driven (ITG) mode in stellarator geometry. In 
tokamak geometry, we compare the results of the new scheme with previous
simulations \cite{Mishchenko_MHD}. In stellarator geometry, we show that the
simulations become feasible for the parameters considered only when the new scheme is
applied. Otherwise, a severe numerical instability develops, caused by the
cancellation problem. 

The structure of the paper is as follows. In Sec.~\ref{Method} the method
suggested is described. In Sec.~\ref{Simulations} the simulations verifying
this method and demonstrating its performance are discussed. The conclusions
are drawn in Sec.~\ref{Conclusions}.
%
%
\section{Pullback mitigation of the cancellation problem} \label{Method}
To derive the mitigation scheme, we deliberately split the magnetic potential
into the `symplectic' and `hamiltonian' parts:
\be
\label{splitting}
A_{\|} = A_{\|}^{\rm(s)} + A_{\|}^{\rm(h)}
\ee
This naming is inspired by Ref.~\cite{Brizard:review}; the precise relation
will become more clear in the following. 
In these notations, the perturbed guiding-center phase-space Lagrangian
\cite{Brizard:review} is
\be
\gamma = q \vc{A}^* \cdot \df \vc{R} + \frac{m}{q} \, \mu \, \df \theta + 
q \, A_{\|}^{\rm(s)} \vc{b} \cdot \df \vc{x}  + 
q \, A_{\|}^{\rm(h)} \vc{b} \cdot \df \vc{x} 
- \left[ \frac{m v_{\|}^2}{2} + \mu B
+ q \phi \right]  \df t \nonumber
\ee
We now perform the Lie transform in such a way that the `hamiltonian part'
$A_{\|}^{\rm(h)}$ contributes to the gyrokinetic Hamiltonian, whereas the
`symplectic part' $A_{\|}^{\rm(s)}$ enters the gyrokinetic symplectic structure
(this explains the naming employed). The resulting gyrokinetic
phase-space Lagrangian is written to first order:
\be
\label{Gamma}
\Gamma = q \vc{A}^* \cdot \df \vc{R} + \frac{m}{q} \mu \, \df \theta + q
\Bgav{A_{\|}^{\rm(s)}} \cdot \df \vc{R} - \left[ \frac{m v_{\|}^2}{2} + \mu B
+ q \Bgav{\phi - v_{\|} A_{\|}^{\rm(h)}} \right]  \df t 
\ee
Here, $\gav{\ldots}$ is the gyro-average, defined as usual. 
The formulation Eq.~(\ref{Gamma}) is neither hamiltonian nor symplectic and will, therefore, be
dubbed the `mixed-variable' formulation, following
Ref.~\cite{Mishchenko_MHD}. 
The corresponding perturbed equations of motion are
\ba
\label{dotR1}
&&{} \dot{\vc{R}}^{(1)} = \frac{\vc{b}}{B_{\|}^*} \times \nabla \Bgav{ \phi - 
  v_{\|} A_{\|}^{\rm(s)} - v_{\|} A_{\|}^{\rm(h)} } - 
\frac{q}{m} \,\gav{A^{\rm(h)}_{\|}} \, \vc{b}^* \\
\label{dotp1}
&&{} \dot{v}_{\|}^{(1)} = \,-\, \frac{q}{m} \, 
\left[ \vc{b}^* \cdot \nabla \Bgav{\phi - v_{\|} A_{\|}^{\rm(h)}} + \pard{}{t}
  \Bgav{A_{\|}^{\rm(s)}} \right] 
 -  \frac{\mu}{m} \, \frac{\vc{b} \times \nabla B}{B_{\|}^*} \cdot \nabla
  \Bgav{A_{\|}^{\rm(s)}}
\ea
For the scheme to work, an equation for $\partial A_{\|}^{\rm(s)} /
\partial t$ is needed. For example, one can follow Ref.~\cite{Mishchenko_MHD}
and use the ideal Ohm's law,  employing the definition:
\be
\label{Ohm}
\pard{}{t}A_{\|}^{\rm(s)} + \vc{b} \cdot \nabla \phi = 0
\ee
This approach appears to be most suitable for Alfv\'enic modes and is utilised
throughout this paper. However, other equations for $\partial A_{\|}^{\rm(s)} /
\partial t$ can be considered, too. Such 
flexibility may be of interest for further optimisation of the simulation
algorithm, but this is beyond the scope of the present paper.  

The zeroth-order gyrocenter characteristics are as usual: 
\be
\label{dotR_p0}
\vc{\dot{R}}^{(0)} = v_{\|} \vc{b}^{*} +
\frac{1}{q B_{\|}^{*}} \vc{b} \times \mu \nabla B  \ , \;\;\;
\dot{v}_{\|}^{(0)} = {}-\, \frac{\mu}{m} \; \vc{b}^{*} \cdot \nabla B 
\ee
Here the following notation has been used:
\ba
&&{} B_{\|}^* = \vc{b} \cdot \vc{\wt{B}}^* \ , \;\;\;
\vc{\wt{B}}^* = \vc{B}^* + \nabla \Bgav{A_{\|}^{\rm(s)}} \times \vc{b} \\
&&{} \vc{B}^*  = \vc{B} + \frac{m v_{\|}}{q} \nabla \times \vc{b} \ , \;\;\;
\vc{b}^* = \vc{B}^* / B_{\|}^*
\ea
The mixed-variable distribution function is solved from the gyrokinetic Vlasov
equation: 
\be
\label{GKE}
\pard{f_{1s}^{\rm(m)}}{t} + 
\dot{\vc{R}}^{(0)} \cdot \pard{f_{1s}^{\rm(m)}}{\vc{R}}  + 
\dot{v}_{\|}^{(0)} \pard{f_{1s}^{\rm(m)}}{v_{\|}} = 
\,-\, \dot{\vc{R}}^{(1)} \cdot \pard{F_{0s}}{\vc{R}}  -
\dot{v}_{\|}^{(1)} \pard{F_{0s}}{v_{\|}} 
\ee 
Here, the index $s = i,e,f$ denotes the particle species (ions, electrons, or
fast ions); $F_{0s}$ is the non-perturbed distribution function (usually a
Maxwellian); the gyrocenter orbits are given by Eqs.~(\ref{dotR1}),
(\ref{dotp1}) and (\ref{dotR_p0}). 
In this paper, we apply the linearised version of the gyrokinetic equation,
but the algorithm, described below, can also be used in nonlinear regime. 

The electrostatic potential and the `hamiltonian part' of the magnetic
potential are found from the gyrokinetic quasineutrality equation and
mixed-variable parallel Ampere's law, respectively: 
\ba
\label{qasi}
&&{} \int \frac{q_i F_{0i}}{T_i} \, (\phi - \gav{\phi}) \, \delta(\vc{R} +
\bm{\rho} - \vc{x}) \, \df^6 Z  
 = \bar{n}_{1i} - \bar{n}_{1e} \\
\label{amp}
&&{} \left( \frac{\beta_i}{\rho_{i}^{2}} +
\frac{\beta_e}{\rho_{e}^{2}} - 
\nabla_{\perp}^{2} \right) A_{\|}^{\rm(h)}
- \nabla_{\perp}^2 A_{\|}^{\rm(s)} 
= \mu_{0} \left( \bar{j}_{\|1i} + \bar{j}_{\|1e} \right) 
\ea
with the usual notations: the mixed-variable gyrocenter density 
$\bar{n}_{1s} = \int \df^{6}Z \, f_{1s}^{\rm(m)} \, \delta(\vc{R} +
\bm{\rho} - \vc{x})$, whose relation to the physical density depends on the 
particular formulation of the gyrokinetic theory used; 
the mixed-variable gyrocenter current
$\bar{j}_{\|1s} = q_{s} \int \df^{6}Z \, f_{1s}^{\rm(m)} \, v_{\|} \,
\delta(\vc{R} + \bm{\rho} - \vc{x})$, related to the physical current by the
pullback; the particle charge $q_s$; the gyrokinetic phase-space volume 
$\df^{6}Z = B_{\|}^{*} \, \df \vc{R} \, \df v_{\|} \, \df \mu \, \df \theta$; 
the thermal gyroradius $\rho_{s} = \sqrt{m_{s}T_{s}}/(eB)$ and 
$\beta_s = \mu_{0} n_{0} T_{s}/B_{0}^{2}$. 

Now, we consider how the mixed-variable formalism can be used in order to
mitigate the cancellation problem. This problem appears in the conventional
hamiltonian formulation of the gyrokinetic theory. In this
formalism, the `parallel velocity' variable is defined as 
\be
v_{\|}^{\rm(h)} = v_{\|}^{\rm(gc)} + \frac{\vc{b}^*}{B} \cdot \nabla \left[ 
\intl^{\theta_{\rm(gc)}} \Big( \psi - \gav{\psi} \Big) \, \df \theta_{\rm(gc)} \right] 
+ \frac{e}{m} \, A_{\|} \ , \;\;\; \psi = \phi - v_{\|}^{\rm(gc)} A_{\|} 
\ee
Here, $v_{\|}^{\rm(gc)}$ is the usual guiding-center parallel velocity and 
$\theta_{\rm(gc)}$ is the guiding-center gyro-phase. 
The cancellation problem can be related to the last term $(e/m) \, A_{\|}$ in
this definition. In contrast, this term is modified and the
cancellation problem is absent in the symplectic formulation, with the
`parallel velocity' 
\be
v_{\|}^{\rm(s)} = v_{\|}^{\rm(gc)} + \frac{\vc{b}^*}{B} \cdot \nabla \left[ 
\intl^{\theta_{\rm(gc)}} \Big( \psi - \gav{\psi} \Big) \, \df \theta_{\rm(gc)} \right] 
+ \frac{e}{m} \, \wt{A}_{\|}
\ , \;\;\; 
\wt{A}_{\|} = A_{\|} - \gav{A_{\|}}
\ee
The mixed-variable formulation is intermediate between the hamiltonian and the
symplectic formulations, with the `parallel velocity' variable defined as 
\be
v_{\|}^{\rm(m)} = v_{\|}^{\rm(gc)} + \frac{\vc{b}^*}{B} \cdot \nabla \left[ 
\intl^{\theta_{\rm(gc)}} \Big( \psi - \gav{\psi} \Big) \, \df \theta_{\rm(gc)} \right] 
+ \frac{e}{m} \, A_{\|}^{\rm(h)} + \frac{e}{m} \, \wt{A}_{\|}^{\rm(s)} \ , \;\;\; 
\wt{A}_{\|}^{\rm(s)} = A_{\|}^{\rm(s)} - \Bgav{A_{\|}^{\rm(s)}}
\ee
The cancellation problem is still present in this formulation but it can be
mitigated by minimising $A_{\|}^{\rm(h)}$, which in contrast to $A_{\|}$ is an
arbitrarily chosen quantity. Note that the mixed-variable formulation
becomes identical to the symplectic when $A_{\|}^{\rm(h)} = 0$.

Invoking the pullback transformation \cite{Brizard:review}, one can express 
the distribution function in the symplectic formulation through the
mixed-variable distribution function as follows:
\be
\label{pullback}
f_{1s}^{\rm(s)} = f_{1s}^{\rm(m)} + \frac{q_s \, \gav{A_{\|}^{\rm(h)}}}{m_s}  \,
\pard{F_{0s}}{v_{\|}} 
\ee
This equation results from the scalar nature of the distribution
function, which implies $f_s^{\rm(s)}[v_{\|}^{\rm(s)}] =
f_s^{\rm(m)}[v_{\|}^{\rm(m)}]$ for the total distribution functions
$f_s^{\rm(s)} = F_{0s} + f_{1s}^{\rm(s)}$ and 
$f_s^{\rm(m)} = F_{0s} + f_{1s}^{\rm(m)}$. 
Here again one sees that the cancellation problem, 
absent in the symplectic formulation, 
can be greatly mitigated by minimising the difference  
$|f_{1s}^{\rm(m)} - f_{1s}^{\rm(s)}|$, or, equivalently, keeping 
the dominant part of the parallel vector potential in its `symplectic part' 
$A_{\|} \approx A_{\|}^{\rm(s)}$ which results in 
\be
A_{\|}^{\rm(h)} \ll A_{\|} 
\ee
In some cases, this can be achieved 
utilising certain ideas about the physical properties of the system under 
consideration, such as 
a particular form of Ohm's law which can be used to determine the
physically dominant part of the magnetic potential \cite{Mishchenko_MHD}. 
An alternative to 
this approach is to numerically accumulate the value of the magnetic potential
in its symplectic part. For this purpose, we can modify the usual algorithm as
follows. 
\begin{enumerate}
\item At the end of each time step, redefine the magnetic potential splitting,
  Eq.~(\ref{splitting}), so that the entire instantaneous value of the parallel
  magnetic potential $A_{\|}(t_i)$ is collected in its `symplectic part':
\be
\label{ic_A}
A_{\|\rm(new)}^{\rm(s)}(t_i) = A_{\|}(t_i) = A_{\|\rm(old)}^{\rm(s)}(t_i) +
A_{\|\rm(old)}^{\rm(h)}(t_i) 
\ee
\item As a consequence of the new splitting, Eq.~(\ref{ic_A}), the
  `hamiltonian' part of the vector potential must be corrected:
\be
\label{Ah_new}
A_{\|\rm(new)}^{\rm(h)}(t_i) = 0
\ee
\item For this modified splitting, the new mixed-variable distribution
  function must coincide with its symplectic-formulation counterpart. The
  symplectic-formulation distribution function is independent on the way of
  splitting and can be found invoking the
  pullback, Eq.~(\ref{pullback}), and using the old values of the
  mixed-variable distribution function and the `hamiltonian' part of the
  parallel vector potential found solving, respectively, the gyrokinetic
  equation~(\ref{GKE}) and Ampere's law, Eq.~(\ref{amp}), at the current time
  step $t_i$: 
\be
\label{ic_f}
f_{1s\rm(new)}^{\rm(m)}(t_i) = f_{1s}^{\rm(s)}(t_i) = 
f_{1s\rm(old)}^{\rm(m)}(t_i) + \frac{q_s \, \gav{A_{\|\rm(old)}^{\rm(h)}(t_i)}}{m_s}  \,
\pard{F_{0s}}{v_{\|}} 
\ee
\item Proceed, explicitly solving the mixed-variable system of equations 
  (\ref{dotR1})-(\ref{amp}) at the next time step $t_i + \Delta t$ in a usual
  way, but using Eqs.~(\ref{ic_A})-(\ref{ic_f}) as the initial conditions.
%
\end{enumerate}
This rearrangement between the symplectic and the hamiltonian components
of the `initial conditions' 
has to be done regularly, i.e.~at each time step. Note that the parallel physics is determined by 
the time derivative of the magnetic potential, whereas the cancellation
problem is proportional to its instantaneous value. 
In our approach, we force the `symplectic part' $A_{\|}^{\rm(s)}$ to be a
dominant contribution to this value. 
The small residual $A_{\|}^{\rm(h)}$ is 
self-consistently computed at each time 
step from the gyrokinetic system of equations in the mixed-variable
formulation, thus guaranteeing correctness of the physical quantity
$\partial A_{\|} / \partial t$, in accordance with the actual dynamics of the
system. This `hamiltonian' correction, being very small, will not lead to a
cancellation problem of any significance. The scheme is not limited to Alfv\'enic 
systems, which obey $E_{\|} \approx 0$, and will work independently of the 
particular physical properties of the system considered. Since the key part of our 
approach, the distribution function transformation Eq.~(\ref{ic_f}), is
directly related to the pullback transform, Ref.~\cite{Brizard:review}, we
call it the `pullback mitigation' of the cancellation problem. 
%
%
\section{Simulations} \label{Simulations}
In this section, we verify the scheme suggested above. We apply the
particle-in-cell code EUTERPE \cite{kornilov04}, a 
non-axisymmetric extension of the GYGLES code
\cite{Mishchenko_tae,Mishchenko_fast,Mishchenko_bulk,Mishchenko_kink}, 
in tokamak and stellarator geometries. 
For consistency, we give here a short description of the numerical scheme used
(also described elsewhere \cite{kornilov04}).

The code solves the gyrokinetic equation using the characteristics
Eqs.~(\ref{dotR1}), (\ref{dotp1}) and (\ref{dotR_p0}). The perturbed fields
$\phi$, $A_{\|}^{\rm(h)}$ and $A_{\|}^{\rm(s)}$ are found numerically solving
the quasineutrality equation (\ref{qasi}), parallel Amp\`ere's law, Eq.~(\ref{amp}), and 
parallel Ohm's law, Eq.~(\ref{Ohm}). Here, the first two equations
(\ref{qasi}) and (\ref{amp}) represent boundary-value problems whereas the
last one, Eq.~(\ref{Ohm}), is an initial value-problem. 
We choose $A_{\|}^{\rm(s)}(t=0) = 0$ as the initial condition for Ohm's law.  
The perturbed part of the distribution function is discretised 
with markers: 
\be
\label{pic0}
f^{\rm(m)}_{1s}(\vc{R},v_{\|},\mu,t) = \sum_{\nu=1}^{N_p} 
w_{s\nu}(t)  
\delta(\vc{R} - \vc{R}_{\nu}) \delta(v_{\|} - v_{\nu \|}) 
\delta(\mu - \mu_{\nu}) \ ,
\ee
where $N_p$ is the number of markers, $(\vc{R}_{\nu},v_{\nu \|},\mu_{\nu})$
are the marker phase space coordinates and $w_{s\nu}$ is the weight of a
marker.
The electrostatic and magnetic potentials are discretized with the
finite-element method (Ritz-Galerkin scheme): 
\be
\label{fin_el_discr}
\phi(\vc{x},t) = \sum_{l=1}^{N_s} \phi_l(t) \Lambda_l(\vc{x}) \ , \;\;\;
A_{\|}^{\rm(h)}(\vc{x},t) = \sum_{l=1}^{N_s} a^{\rm(h)}_l(t) \Lambda_l(\vc{x}) \ ,
\;\;\; 
A_{\|}^{\rm(s)}(\vc{x},t) = \sum_{l=1}^{N_s} a^{\rm(s)}_l(t) \Lambda_l(\vc{x}) \ ,
\ee
where $\Lambda_l(\vc{x})$ are the finite elements (tensor product of B
splines \cite{de_Boor,Hoellig}); $N_s$ is the total
number of the finite elements; $\phi_l$, $a^{\rm(h)}_l$ and $a^{\rm(s)}_l$ are
the spline coefficients. 
In this formulation, the gyrokinetic equation (\ref{GKE}) corresponds to the evolution of
the marker weights $w_{s\nu}(t)$ and the parallel Ohm's law, Eq.~(\ref{Ohm}),
translates into the evolution of the spline coefficients $a^{\rm(s)}_l(t)$.  
A more detailed description of the discretization procedure can
be found in Refs.~\cite{Fivaz_1998,Mishchenko1,Mishchenko2,Mishchenko3,Hatzky_2007}. 
We apply the so-called phase factor transform \cite{Fivaz_1998} to all
perturbed quantities in the code. 
The integrals over the gyro-angle 
are approximated with an N-point discrete
sum \cite{Lee,Hatzky02,Mishchenko3}. 
%
The cancellation problem
\cite{Chen_Parker_2001,Mishchenko1}, which in the mixed-variable 
formulation is related only to the correction $A_{\|}^{\rm(h)}$ 
of the parallel magnetic potential [see Eq.~(\ref{amp})], 
is solved using the iterative scheme introduced in
Refs.~\cite{Chen_Parker_2003,Hatzky_2007}. 


First, we consider a TAE in a tokamak configuration with a large aspect ratio
and a circular cross section, the minor radius  
$r_{\rm a} = 1$~{\rm m}, the major radius $R_0 = 10$~{\rm m}, the magnetic
field on axis $B_0 = 3$~{T}, and the safety factor profile 
$q(r) =  1.71 + 0.16 \, (r / r_{\rm a})^2$ (here, $r$ is the small radius). 
The background plasma profiles (corresponding to Maxwellian unperturbed
distribution functions) are chosen 
to be flat with the ion (hydrogen) and electron densities 
$n_{i} = n_e = 2 \times 10^{19}~{\rm m^{-3}}$, and flat temperatures 
$T_i = T_e = 1$~{\rm keV}. 
A Maxwellian is also chosen for the unperturbed
distribution function of the fast particles (deuterium ions). The
fast particle temperature $T_f$ is flat and the fast particle density is given
by the expression: 
\be
\label{denf_eq}
n_{f}(s_{\rm pol}) = n_{0f} \exp\left[ {}- \frac{\Delta_{{\rm n}f}}{L_{{\rm n}f}} \, 
{\rm tanh}\left(\frac{s_{\rm pol} - s_{{\rm n}f}}{\Delta_{{\rm n}f}}\right) \right] 
\ee
with $s_{\rm pol}$ being the square root of the normalised poloidal flux, $s_{{\rm n}f} = 0.5$ the
position of the maximal value of  
$\kappa_{{\rm n}f} = |\nabla n_f| / n_f$, $n_{0f} = 0.75 \times
10^{17}~{\rm m^{-3}}$ the fast particle density at $s_{\rm pol} = s_{{\rm n}f}$,
$\Delta_{{\rm n}f} = 0.2$ the characteristic width of the density profile, and $L_{{\rm
    n}f} = 0.3$ determining the strength of the fast particle density gradient. 

In this configuration, we simulate the TAE with
the toroidal mode number $n = \,-\,2$ and the dominant poloidal harmonics $m =
3$ and $m = 4$. This case is particularly difficult, since it corresponds to
the so-called MHD limit at low perpendicular mode numbers, where the
cancellation problem is most severe. It has been considered in
Ref.~\cite{Mishchenko_MHD} and is used here again to verify the pullback
mitigation scheme.

In Figs.~\ref{n2_tf_w} and \ref{n2_tf_g}, the frequency and the growth rate of
the TAE are shown, respectively, as functions of the
fast-ion temperature. The result obtained with the pullback mitigation is
compared with the MHD mitigation \cite{Mishchenko_MHD} simulations and with
the electron-fluid scheme \cite{MCole_kink} simulations (the electron-fluid
scheme is similar to Refs.~\cite{ChenandParker,Lin_and_Chen}). One
sees that the agreement is excellent. Interestingly, the usual physical
stabilization effect \cite{Gorelenkov_FOW99} due to the finite orbit width is
rather weak here, in contrast to the moderate mode number simulations
presented in Ref.~\cite{Mishchenko_fast}. Such a scaling is to be expected at the
small dominant perpendicular mode numbers, considered in the present simulations.

We continue our numerical experiments with the pullback mitigation scheme in
stellarator geometry. A magnetic geometry similar to the Large Helical Device
(LHD) \cite{lhd} is considered. The plasma is chosen to have 
$\beta_* = \mu_0 n_* T_* / B_*^2 = 0.0085$. Here $n_*$ is the plasma density
averaged over the entire plasma volume, $T_* = T_e(s = 0.5)$ with $s$ being
the normalised toroidal flux and $B_* = B(s=0, \zeta=0)$ with $\zeta$ being
the toroidal angle. The  plasma size is
determined by the parameter $L_{\rm x} = 900$ which is approximately the 
ratio $L_{\rm x} \approx 2.2 \, r_a / \rho_s$ with $r_a$ being the average minor radius of the
non-axisymmetric device and $\rho_s = \sqrt{m_i T_*}/(e B_*)$ the
characteristic ion sound
gyroradius. Note that simulations at large values of $L_{\rm x}$ 
(small values of $\rho_* = \rho_s / r_a$) are particularly challenging since
one can show that the cancellation problem scales as $\beta_* L_{\rm x}^2$. 
The plasma density and temperatures profiles are defined as the
functions of the normalised toroidal flux according to the expressions:
\ba
\label{den_eq}
&&{} n_{(i,e)}(s) = n_{0} \exp\left[ {}- \frac{\Delta_{{\rm n}}}{L_{{\rm n}}} \, 
{\rm tanh}\left(\frac{s - s_0}{\Delta_{{\rm n}}}\right) \right] \\
&&{} T_{(i,e)}(s) = T_{0} \exp\left[ {}- \frac{\Delta_{{\rm T}(i,e)}}{L_{{\rm T}(i,e)}} \, 
{\rm tanh}\left(\frac{s - s_0}{\Delta_{{\rm T}(i,e)}}\right) \right] 
\ea
with $s_0 = 0.5$, $1/L_{{\rm n}} = 1.5$, 
$\Delta_{{\rm n}} = 0.2$, $1/L_{{\rm T}i} = 3.5$, $\Delta_{{\rm T}i} = 0.2$, 
$1/L_{{\rm T}e} = 3.0$, and $\Delta_{{\rm T}e} = 0.2$. The parameters $n_0$
and $T_0$ are determined by the plasma size $L_{\rm x}$ and its pressure 
$\beta_*$.

We consider an electromagnetic mode centered in the Fourier space around the
poloidal mode number $m_0 = \,-\,35$ and the toroidal mode number 
$n_0 = \,-\,23$. This mode can be destabilised by finite ion temperature
gradient, becoming the electromagnetic ITG mode. `Electromagnetic' implies
here that the gyrokinetic electromagnetic system of equations will be solved
in the simulations at plasma $\beta$ exceeding 
the hydrogen electron-to-ion mass ratio. 

We start our simulations using the standard cancellation scheme
\cite{Chen_Parker_2003,Hatzky_2007}. 
In Fig.~\ref{lhd_mhd0itg_vals}, the time evolution of the mode is
shown. One sees that it becomes strongly unstable within a few time steps. The
Fourier spectrum is shown in Fig.~\ref{lhd_mhd0itg_pspect}: it is
completely dominated by the noise at the edge of the Fourier window, caused by
the cancellation problem. Finally, the radial pattern of the
Fourier-decomposed quantity $\phi(s,\theta,\zeta_0)$ is shown in
Fig.~\ref{lhd_mhd0itg_modes}. Here, $\phi$ is the electrostatic
potential, $s$ is the normalised toroidal flux, $\theta$ is the poloidal angle
and $\zeta_0 = 0$ is the particular (fixed) toroidal angle. Different lines
correspond in Fig.~\ref{lhd_mhd0itg_modes} to different poloidal
harmonics of $\phi(s,\theta,\zeta_0)$. 

The cancellation problem can in stellarator geometry be somewhat alleviated with
the MHD mitigation scheme, described in Ref.~\cite{Mishchenko_MHD}.  
In Fig.~\ref{lhd_mhd1itg_vals}, the time evolution is shown in the case
when the MHD mitigation scheme is applied. One sees that the numerical
instability sets in later in this case, but it is still present and disrupts the
simulation. In Fig.~\ref{lhd_mhd1itg_pspect}, the Fourier spectrum of the
mode is shown at the time of the instability onset, showing both the physical
mode, still visible in the center of the Fourier window, accompanied by the
wide-band noise signal appearing around the physical mode. A few time steps
later, the noise dominates and the simulation dies. In
Fig.~\ref{lhd_mhd1itg_modes}, 
the radial pattern of the Fourier-decomposed $\phi(s,\theta,\zeta_0)$ is
shown at the onset of the numerical instability. Again, the physical mode is
still there and coexists with the distinct structure at the edge. This
structure is caused by the cancellation problem and dominates the simulation
after a few more time steps. 

Finally, we describe simulations using the pullback mitigation. 
In Fig.~\ref{lhd_emitg_vals}, the time evolution of the
electromagnetic ITG is shown. One sees now that the mode develops in a clean
physical way. 
All parameters here are identical to those used for Figs.~\ref{lhd_mhd0itg_vals} and
\ref{lhd_mhd1itg_vals}. The only difference is that the pullback
mitigation has been switched on here. In Fig.~\ref{lhd_emitg_pspect}, the
Fourier spectrum is shown. One sees that the mode is numerically clean
(cf.~the structure with Figs.~\ref{lhd_mhd0itg_pspect} or
\ref{lhd_mhd1itg_pspect}) and  shows a ballooning-like structure,
indicating importance of toroidicity 
in the case considered.  In Figs.~\ref{lhd_emitg_modes} and 
\ref{lhd_emitg_plpot}, 
the radial structure of the Fourier-decomposed $\phi(s,\theta,\zeta_0)$ and its poloidal
cross section are shown. 
One sees that the mode appears to be physical and ballooning-like in all these
representations.  
%
%
\section{Conclusions}  \label{Conclusions}
In this paper, we have further developed the mixed-variable gyrokinetic
formalism. In our previous work \cite{Mishchenko_MHD}, the variables were
determined by a particular form of Ohm's law. In the more general formulation
presented here, such a limitation can be relaxed. 

Using the mixed-variable formulation, we have proposed a new 
algorithm 
which can strongly mitigate the cancellation problem: pullback mitigation. The
scheme follows the observation that the parallel dynamics is determined by the
time derivative of 
the parallel magnetic potential whereas the cancellation problem is
proportional to its value. Accumulating this value into the `symplectic' part of
the parallel magnetic potential makes it possible to minimise all terms 
that are relevant for the cancellation problem. Such a correction 
occurs at each time step in the course of a simulation, guaranteeing the
smallness of the `hamiltonian' residual. Note that pullback mitigation is
not limited to the PIC framework. 

We have verified the pullback mitigation approach in tokamak and
stellarator geometries. In tokamak geometry, a Toroidal Alfv\'en Eigenmode with small mode 
numbers (the so-called MHD limit) has been simulated and compared with previous results. Very good
agreement has been found. In stellarator geometry, the electromagnetic ITG
mode has been simulated at a realistic $\rho_*$. It has been shown that the
simulation becomes feasible, for the parameters considered, only when the
pullback mitigation is used. 

We believe that the approaches suggested in this paper and in
Ref.~\cite{Mishchenko_MHD} will greatly facilitate the electromagnetic
simulations, both in tokamak and stellarator geometries. While only linear
simulations have been considered here, our methods should also work 
in nonlinear regimes. We leave detailed study for future work. 
%
%
\qquad \\
\qquad \\
\qquad \\
{\bf ACKNOWLEDGEMENTS}

We acknowledge the support of Per Helander for this work. We thank J\"urgen
N\"uhrenberg for carefully reading the manuscript. Discussions on the
cancellation problem during the workshop on ``Modelling kinetic aspects of
Global MHD Modes'' organised by Egbert Westerhof at the Lorentz Center in
Leiden were very helpful. In particular, we acknowledge remarks by Arthur
Peeters which provided an initial impulse for this paper. This work was carried out
using the HELIOS supercomputer system at Computational Simulation Centre of
International Fusion Energy Research Centre (IFERC-CSC), Aomori, Japan, under
the Broader Approach collaboration between Euratom and Japan, implemented by
Fusion for Energy and JAEA. Also, some simulations have been performed on the
local cluster in Greifswald, where support of Henry Leyh is appreciated. The
project has received funding from the Euratom research and training program 2014-2018. 
%
%
%

\begin{thebibliography}{10}

\bibitem{Chen_Parker_2001}
Y. Chen and S. Parker, Phys. Plasmas {\bf 8},  2095  (2001).

\bibitem{Mishchenko1}
A. Mishchenko, R. Hatzky, and A. {K\"onies}, Phys. Plasmas {\bf 11},  5480
  (2004).

\bibitem{Chen_Parker_2003}
Y. Chen and S. Parker, J. Comp. Phys {\bf 189},  463  (2003).

\bibitem{Mishchenko2}
A. Mishchenko, A. {K\"onies}, and R. Hatzky,  in {\em Proc.\ of the Joint
  Varenna-Lausanne International Workshop} ({Societ\`a} Italiana di Fisica,
  Bologna, 2004).

\bibitem{Hatzky_2007}
R. Hatzky, A. {K\"onies}, and A. Mishchenko, J. Comp. Phys. {\bf 225},  568
  (2007).

\bibitem{GYRO}
J. Candy and R.~E. Waltz, J. Comp. Phys. {\bf 186},  545  (2003).

\bibitem{Mishchenko_MHD}
{A.~Mishchenko, M.~Cole, R.~Kleiber, and A.~K\"onies}, Phys. Plasmas {\bf 21},
052113 (2014).

\bibitem{Brizard:review}
A.~J. Brizard and T.~S. Hahm, Reviews of Modern Physics {\bf 79},  421  (2007).

\bibitem{kornilov04}
{V.~Kornilov, R.~Kleiber, R.~Hatzky, L.~Villard, and G.~Jost}, Phys. Plasmas
  {\bf 11},  3196  (2004).

\bibitem{Mishchenko_tae}
A. Mishchenko, R. Hatzky, and A. {K\"onies}, Phys. Plasmas {\bf 15},  112106
  (2008).

\bibitem{Mishchenko_fast}
A. Mishchenko, A. {K\"onies}, and R. Hatzky, Phys. Plasmas {\bf 16},  082105
  (2009).

\bibitem{Mishchenko_bulk}
A. Mishchenko, A. {K\"onies}, and R. Hatzky, Phys. Plasmas {\bf 18},  012504
  (2011).

\bibitem{Mishchenko_kink}
A. Mishchenko and A. Zocco, Phys. Plasmas {\bf 19},  122104  (2012).

\bibitem{de_Boor}
C. de~Boor, {\em A Practical Guide to Splines} (Springer-Verlag, New York,
  1978).

\bibitem{Hoellig}
K. {H\"ollig}, {\em Finite Element Methods with B-Splines} (Society for
  Industrial and Applied Mathematics, Philadelphia, 2003).

\bibitem{Fivaz_1998}
{M.~Fivaz, S.~Brunner, G.~de~Ridder, O.~Sauter, T.~M.~Tran, J.~Vaclavik,
  L.~Villard, and K.~Appert}, Comp. Phys. Commun. {\bf 111},  27  (1998).

\bibitem{Mishchenko3}
A. Mishchenko, A. {K\"onies}, and R. Hatzky, Phys. Plasmas {\bf 12},  062305
  (2005).

\bibitem{Lee}
W.~W. Lee, J. Comp. Phys. {\bf 72},  243  (1987).

\bibitem{Hatzky02}
{R.~Hatzky, T.~M.~Tran, A.~{K\"onies}, R.~Kleiber, and S.~J.~Allfrey}, Phys.
  Plasmas {\bf 9},  898  (2002).

\bibitem{MCole_kink}
{M.~Cole, A.~Mishchenko, A.~K\"onies, R.~Kleiber, and M.~Borchardt}, Phys.
  Plasmas  (accepted for publication in Phys.~Plasmas).

\bibitem{ChenandParker}
Y. Chen and S. Parker, Phys. Plasmas {\bf 8},  441  (2001).

\bibitem{Lin_and_Chen}
Z. Lin and L. Chen, Phys. Plasmas {\bf 8},  1447  (2001).

\bibitem{Gorelenkov_FOW99}
N.~N. Gorelenkov, C.~Z. Cheng, and G.~Y. Fu, Phys. Plasmas {\bf 6},  2802
  (1999).

\bibitem{lhd}
{O.~Motojima, N.~Ohyabu, A.~Komori, O.~Kaneko, S.~Masuzaki, A.~Ejiri, M.~Emoto,
  H.~Funaba, M.~Goto, K.~Ida, H.~Idei, S.~Inagaki, N.~Inoue, S.~Kado, S.~Kubo,
  R.~Kumazawa, T.~Minami, J.~Miyazawa, T.~Morisaki, S.~Morita, S.~Murakami,
  S.~Muto, T.~Mutoh, Y.~Nagayama, Y.~Nakamura, H.~Nakanishi, K.~Narihara,
  K.~Nishimura, N.~Noda, T.~Kobuchi, S.~Ohdachi, Y.~Oka, M.~Osakabe, T.~Ozaki,
  B.~J.~Peterson, A.~Sagara, S.~Sakakibara, R.~Sakamoto, H.~Sasao, M.~Sasao,
  K.~Sato, M.~Sato, T.~Seki, T.~Shimozuma, M.~Shoji, H.~Suzuki, Y.~Takeiri,
  K.~Tanaka, K.~Toi, T.~Tokuzawa, K.~Tsumori, K.~Tsuzuki, I.~Yamada,
  S.~Yamaguchi, M.~Yokoyama, K.~Y.~Watanabe, T.~Watari, Y.~Hamada, K.~Matsuoka,
  K.~Murai, K.~Ohkubo, I.~Ohtake, M.~Okamoto, S.~Satoh, T.~Satow, S.~Sudo,
  S.~Tanahashi, K.~Yamazaki, M.~Fujiwara, and A.~Iiyoshi}, Nucl. Fusion {\bf
  43},  1674  (2003).

\end{thebibliography}

%
\graph{n2_tf_w}{The frequency of the $n=\,-\,2$ TAE mode obtained with the pullback
mitigation scheme compared with the MHD-mitigation \cite{Mishchenko_MHD}.}
%
\graph{n2_tf_g}{Growth rate of the $n=\,-\,2$ TAE mode obtained with the pullback
mitigation scheme compared with the MHD-mitigation \cite{Mishchenko_MHD} and
fluid-electron \cite{MCole_kink} schemes.} 
\graph{lhd_mhd0itg_vals}{The time evolution of the electrostatic
  potential in LHD-like geometry is shown for the case of the standard
  cancellation scheme \cite{Chen_Parker_2003,Hatzky_2007} applied. 
  The time step is $\omega_{{\rm c}i}\,\Delta t = 0.5$. The time evolution is
  measured at different flux surfaces indicated on the figure. The potential
  is calculated at the toroidal angle $\zeta = 0$ and the poloidal angle
  $\theta = 0$. One sees that a severe numerical instability develops in this
  case within a few time steps.} 
\graph{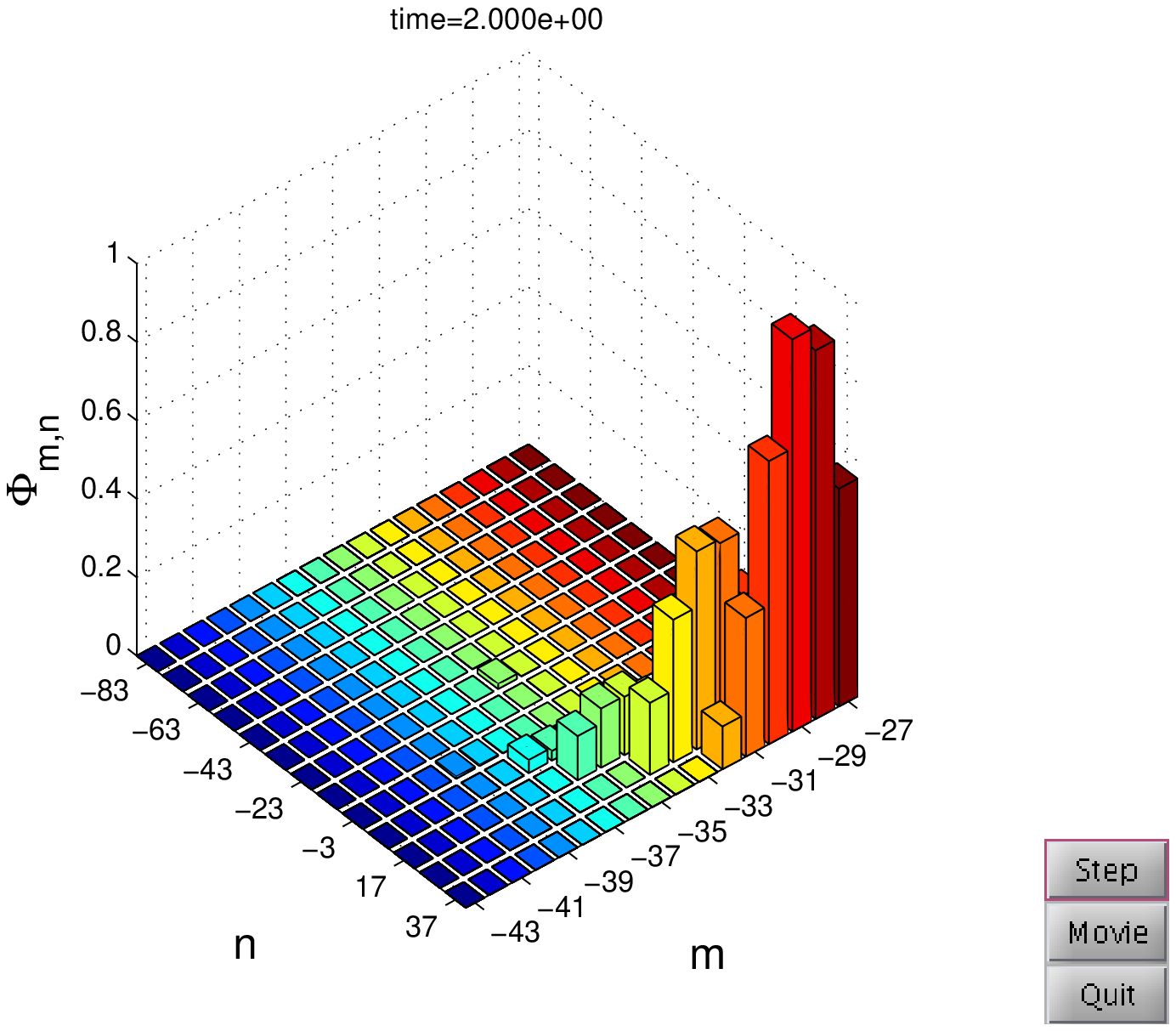}{The Fourier window of the stellarator
  simulations is dominated by the modes at the edge within few time steps. 
  The figure shown corresponds to the fourth time step. The standard 
  cancellation scheme
  \cite{Chen_Parker_2003,Hatzky_2007} is used.}
\graph{lhd_mhd0itg_modes}{The radial pattern developing due to the severe
  numerical instability caused by the cancellation problem. The figure shown
  corresponds to the fourth time step. The standard cancellation scheme
  \cite{Chen_Parker_2003,Hatzky_2007} is used.} 
\graph{lhd_mhd1itg_vals}{The time evolution of the electrostatic
  potential in the LHD-like configuration is shown for the case of the
  MHD-cancellation approach \cite{Mishchenko_MHD} applied. The numerical
  instability is mitigated using this approach, but can not be completely cured.}
\graph{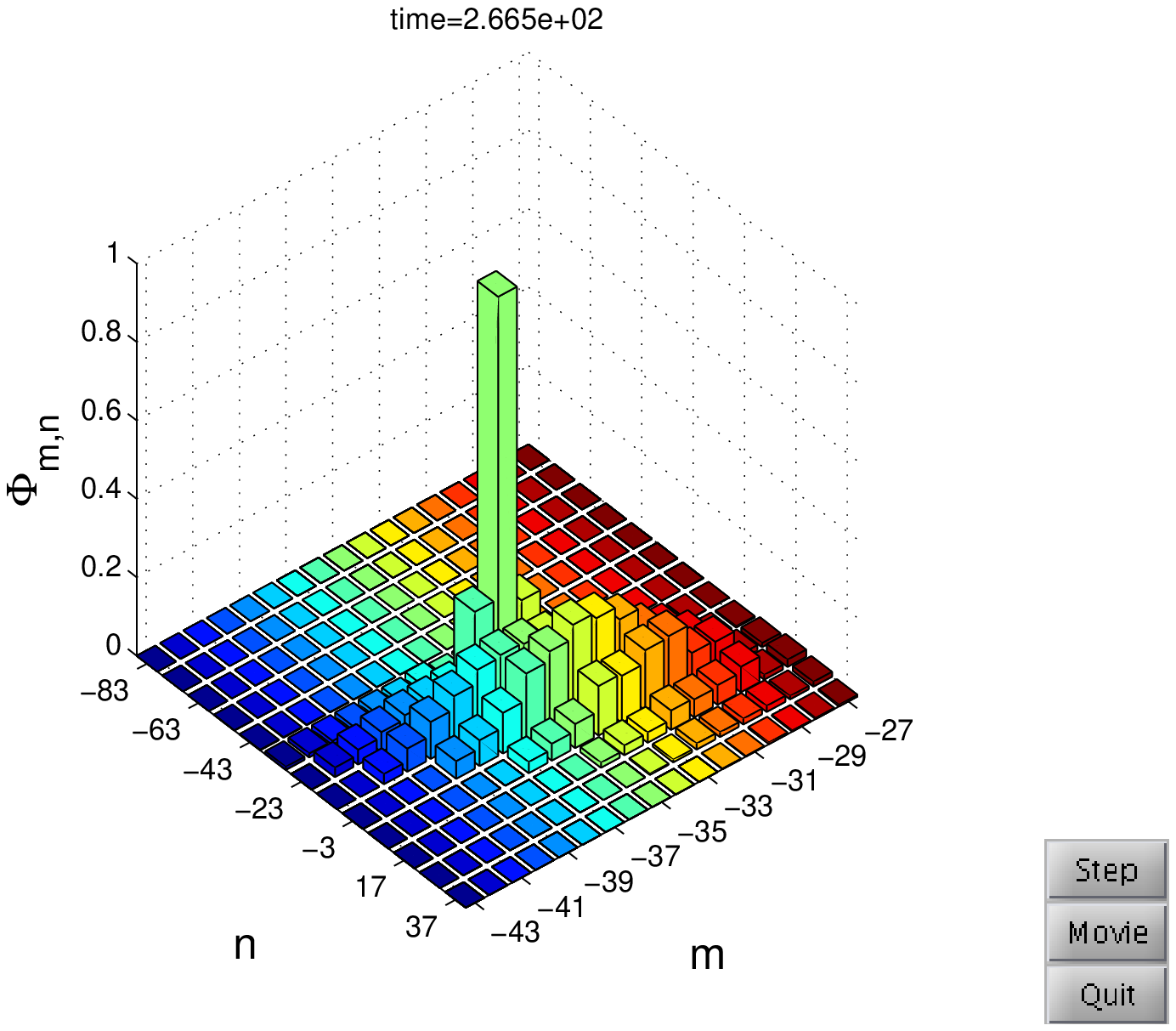}{Onset of the numerical instability from
  Fig.~\ref{lhd_mhd1itg_vals}, here shown in the Fourier window. The
  physical mode can still be seen in the center of the window. In addition,
  one sees a wide-band noise signal appearing. A few time steps later this part
  of the spectrum will dominate the mode completely: numerical
  instability develops.} 
\graph{lhd_mhd1itg_modes}{The radial structure developing during the
  numerical instability, see Fig.~\ref{lhd_mhd1itg_vals}, caused by the
  cancellation problem. Onset of the numerical instability is shown. One sees
  both the physical mode still surviving and the noisy structure at the edge
  growing. A few time steps later, this noisy edge structure will completely
  dominate the radial pattern.} 
\graph{lhd_emitg_vals}{Numerically clean evolution of the physical mode 
  obtained using the pullback mitigation approach. One sees that the mode
  (electromagnetic ITG) grows and has a finite frequency. All the numerical
  parameters coincide with the parameters used for
  Fig.~\ref{lhd_mhd0itg_vals}. The only difference is the pullback
  mitigation, applied here.} 
\graph{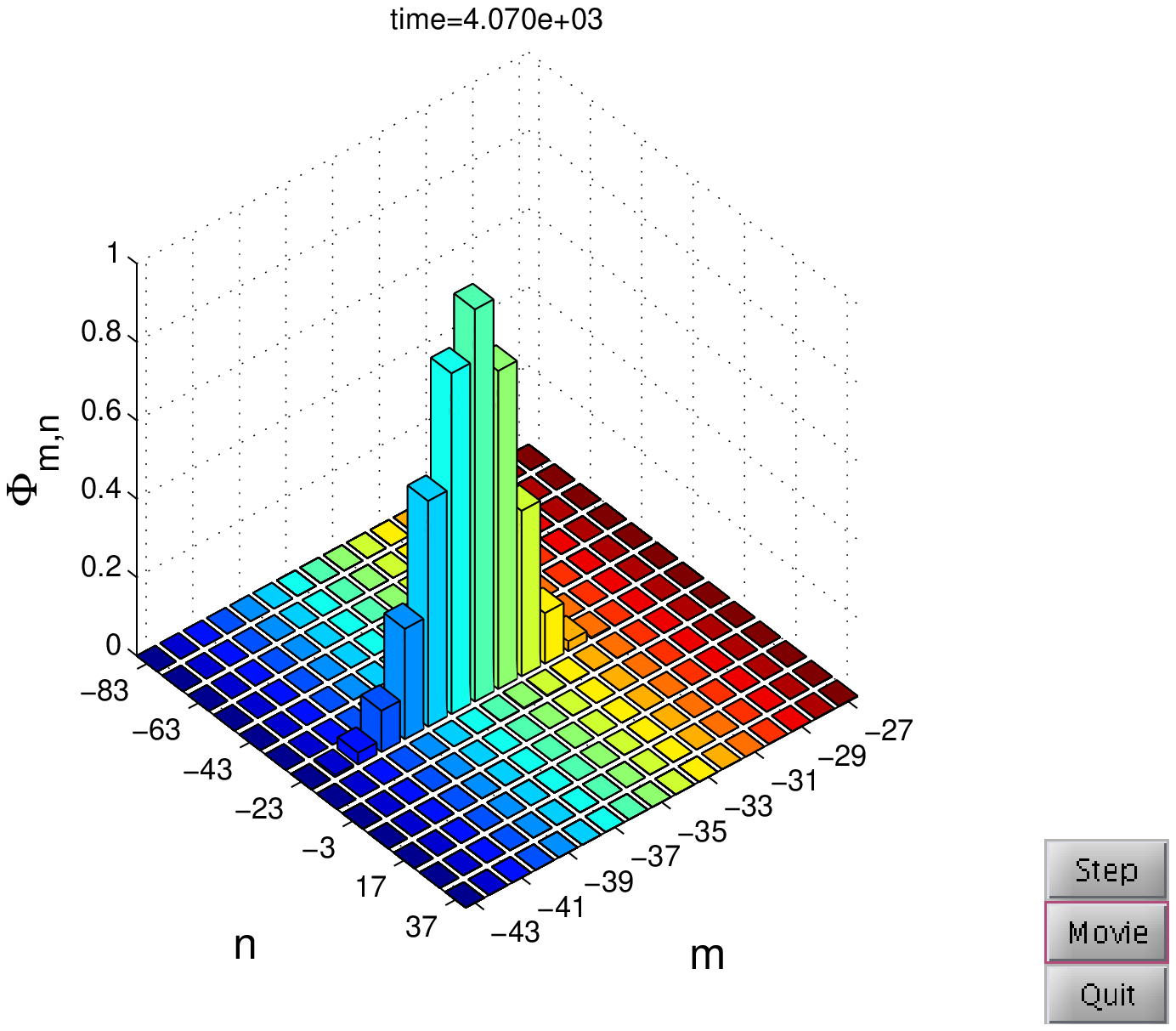}{The physical, numerically clean Fourier spectrum
  of the electromagnetic ITG mode in LHD-like geometry. 
  All the numerical parameters coincide with the parameters used for
  Fig.~\ref{lhd_mhd0itg_pspect}. The only difference is the pullback
  mitigation, applied here.} 
\graph{lhd_emitg_modes}{The radial pattern of the electromagnetic ITG
  mode. All the numerical parameters coincide with
  the parameters used for Fig.~\ref{lhd_mhd0itg_modes}. The only
  difference is the pullback mitigation, applied here.}
\graph{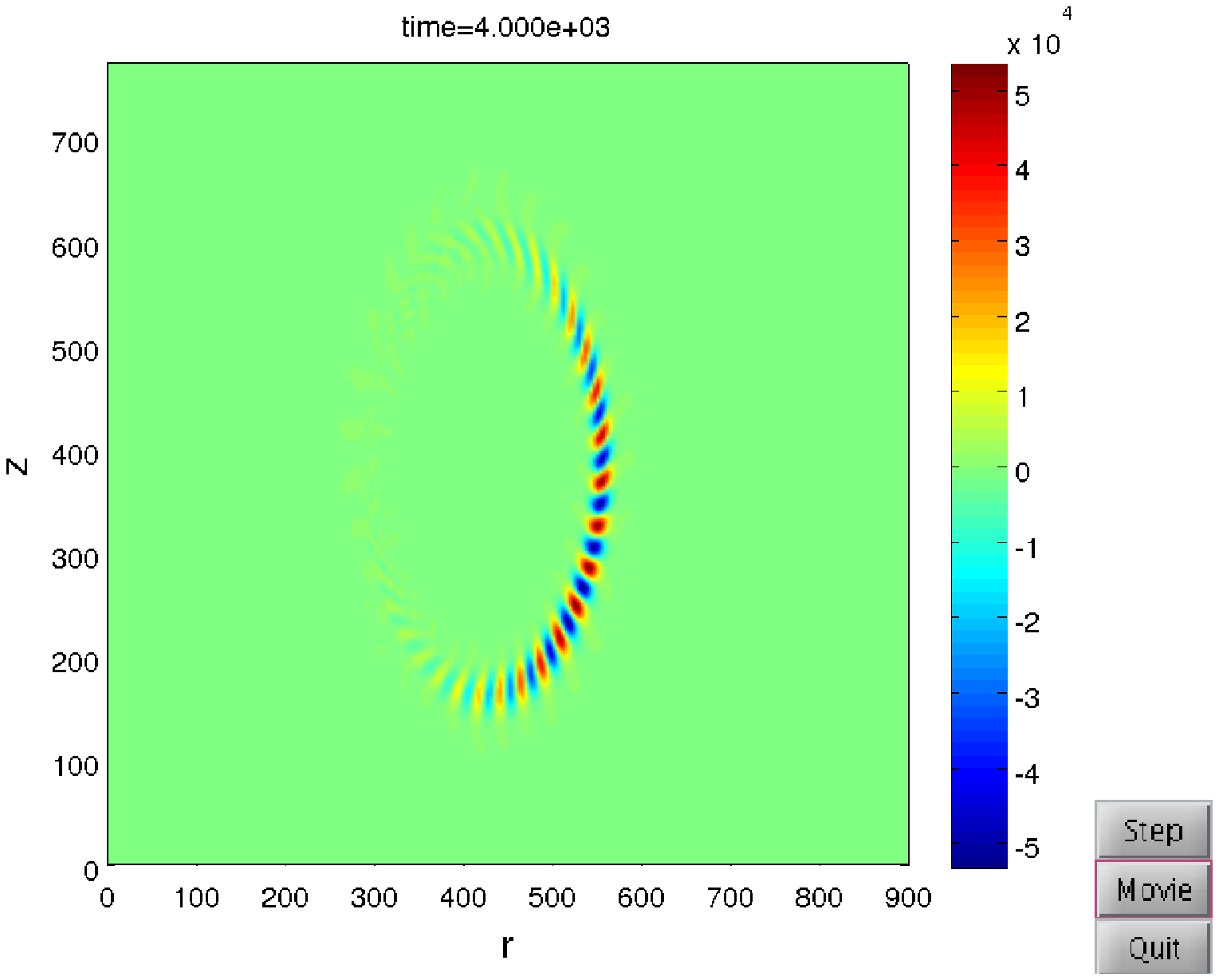}{Electromagnetic ITG mode structure shown as a
  poloidal cross section at the toroidal angle $\zeta = 0$. Pullback
  mitigation of the cancellation problem has been applied.} 
%
%
%
\end{document}